\documentclass[12pt]{article}
\usepackage{graphicx}
\begin{document}
\begin{center}
{\Large {\bf Heavy meson three body decay: \\
Three decades of Dalitz plot amplitude analysis\footnote{To appear at Alberto Santoro Festschrift}.}} \\
\vspace{4mm}
{\bf Ignacio Bediaga\\ Centro Brasileiro de Pesquisas Físicas - CBPF}\\
\vspace{4mm}
{ bediaga@cbpf.br} \\
\end{center}
\vspace*{0.3cm}
\begin{abstract}
{  Three body heavy meson decays  allows  one to extract  important  
quantities   for flavour physic, however the phenomenological   analysis has some  
open problems to be solved  in order to make possible  to extract physical parameters with
precision.  
After many years and some  experience accumulated, we have learned new 
features and we expect to learn much more from data coming  mainly from B three body decays. 
In this paper we  address these aspects and we present a proposal to modify the Dalitz
fit amplitude method in order to  allow further studies. }
\end{abstract}
\vspace*{0.3cm}

\section{Introduction}

  Three body heavy meson decay analysis have been  treated as a coherent sum of individual   
 two body resonances amplitudes plus a non-resonant component, each amplitude with a strong final state interaction phase (FSI).  
 This simple configuration based in the dominance of the  quasi two body component, was proposed in  the early eighties by 
Mark collaboration \cite{mark}. Important changes were made to improve this approach  during these already three decades, 
however  remains the basic  structure  of the original model. This analysis method  have been working well to represent  
low statistics charm three body decay but have been presenting some problems to   represent the equivalent three body beauty  decays.

The  basic concept used by Mark collaboration in its  initial three body analysis  was
inspired  in the amplitude analysis used for $\pi$ - nucleon and 
$K $ - nucleon interaction giving  three or four final state
\cite{Jongejans1,Jongejans2,gidal, michael,Jongejans3,slac}  \footnote{ most of these  experimental studies 
were to search for axial-vector particle properties of a $J^p =  1^+$ mesons decaying   
in three pseudo-scalar mesons \cite{santoro}}.  Nevertheless, there is an important difference between 
hadron hadron interaction and heavy meson decay: while the last one use a full  coherence sum, 
in  hadronic collisions this quantum effect had been used  in a more  restrictively form, assuming 
partially the   coherence between  different two body resonances amplitudes present in the same phase space region. 

Recently, for  some charm meson three body  decay   analysis \cite{focus3pi, focusk2pi}, a significative change  
was partially introduced by  FOCUS collaboration. To take in account   interferences among resonances with the same 
quantum number, FOCUS used the two body unitary K-matrix approach  to represent  the s-wave component.  

The K-matrix in Dalits analysis and the usual approach of   coherent  sum of resonances,
reflected two different treatment  for particle  interferences   well characterized
by Azimov   in a recent paper\cite{azimov}. According to this author,  interference
between particles with the same quantum number would be classified as 
"direct interference", otherwise it would said   to be by  "re-scattering
interference". Among other differences between they two, the locality is one 
of more important characteristic pointed out in this paper. Direct interferences
have a large phase space range, while  re-scattering interference between two 
resonances is  limited at the region  where resonances have intercepting   phase space.

However, if for one side  the K-matrix approach tried to treat correctly  interferences between resonances belonging to 
the same partial wave, on the other side    neglecting  possible re-scattering interferences between  particles with different quantum numbers, 
that is a singularity triangle type diagram \cite{anisovich_sarantsev}. This approximation seems to work for $D^+_s\to \pi^-
\pi^+\pi^+$ decay \cite{antimo}, but not for $D^+ \to K^- \pi^+ \pi^+$. In fact the  $K \pi$  phase motion of the last decay,  extracted with   
 Partial Wave Amplitude (PWA) analysis first provided  by E791 collaboration and then by CLEO and FOCUS \cite{brian,pwa_focus,pwa_cleo}, 
shown a  distribution  substantially different from the one observed in scalar partial wave $K \pi$ elastic scattering by
 LASS  collaboration \cite{LASS} and also  different from the recently PWA analysis performed for the  semi-leptonic  
$D^+ \to K^- \pi^+ e^+ \nu_e$ decays \cite{semi_babar} \footnote{ Both, the  semi-letonic and the  $K \pi$ elastic scattering scalar phase motion 
agree each other in a clear verification   of the Watson theorem \cite{Watson}}. However  the same analysis for the $D^+_s\to \pi^-
\pi^+\pi^+$ decays seems to have similar phase motion behaviour than  the one obtained by  $\pi^-\pi^+$ elastic scattering   \cite{antimo}.

Since both kinds of interferences, with  different behaviour, can be  present in three
body heavy meson decay, a complete amplitude analysis  has to take in account the existence 
of both effects and also provide  from the fit parameters, the relative contribution   of each effect 
for each intermediary state.

\section{  Isobaric model and experimental results.}

 With the increase of  statistics, more resonances with different spins  started to
contribute   significantly in  three body D decays, Argus collaboration introduced 
a generalization of the isobar model \cite{argus}:   each possible resonance 
amplitude is represented by a Breit-Wigner  multiplied by angular
distribution function associated with the spin of the resonance. The Non-Resonant is parametrized as a
constant.   The various contributions are combined in a coherent sum with complex 
coefficients that are  extracted from maximum likelihood fits to the data. The
absolute value of the coefficients are related to the relative fraction of each
contribution and the phases takes into account final state interaction (FSI) between the
resonance and the bachelor pion. This phase is considered constant because it
depends only on  the total energy of the system, i.e. the heavy meson mass.

With this simple model plus  the E791 method \cite{sigma791} to  determine the mass
and the width of some not yet  well  defined resonances, most of three body charm
meson decay were successfully described \footnote{ There is at least one  important
exception, the $  D^0 \to K^0_s \pi^-\pi^+ $. Because CP violation studies 
this decay  have been analysed  by the B-factories with a large samples and  also
many problems \cite{BelleD0, BabarD0}}. On the other hand,  B charmless three body
decay scenario  is not successfully describe by the use of the Isobar model. The most important
problem is associated  with the non-resonant (NR) contribution.  Differently from
charm decays, the NR contribution  has playing  a significant role in $B$ meson studies.  At a very
early stage of the $B$ factory analysis programme the `constant amplitude' 
description of the NR term was abandoned  and replaced by alternative approaches. 
When studying the decay $B^+ \to K^+ \pi^+ \pi^-$ the Belle collaboration proposed
an empirical exponential function to represent the NR contribution \cite{bellek2pi}.
BABAR  instead, used for the same decay  mode~\cite{babark2pi} a phase and magnitude
parametrization obtained by the LASS experiment~\cite{LASS} for S-wave $K\pi$
elastic scattering.

 The empirical function proposed by Belle~\cite{bellek2pi} has two terms, 
one for each Dalitz plot variable 
($s_{13}\equiv M^2(K^+\pi^-)$ and $s_{23}\equiv M^2(\pi^+\pi^-)$), 
and is given by: 
\begin{eqnarray}
  {\cal A}_{nr}(K^+\pi^+\pi^-)=a_1^{nr}e^{-\alpha s_{13}}
  e^{i\delta_1^{nr}}+a_2^{nr}e^{-\alpha s_{23}}e^{i\delta_2^{nr}}\;,
  \label{eq:anr}
\end{eqnarray}
with five free parameters: one magnitude and one  phase for each  of these
terms and  the slope of the exponential function $\alpha$. 
The fit results exhibit three noteworthy and correlated characteristics: 
i) a large NR fraction of around 34\%,
ii) a phase difference between $\delta_1^{nr}$ and $\delta_2^{nr}$ around
$180^\circ$, 
iii) a sum of fractions from all components, resonant plus NR  of around 150\%. 
In an amplitude analysis the sum of fractions can of course be bigger or less 
than 100\% and this is not inconsistent with the unitary concept \footnote{ 
  The concept of fraction in a Dalitz plot analysis is directly related to the
  branching ratio  of a particular amplitude. It can be  classically
  associated to that fraction of the total number of  events in an incoherent
  sum of amplitudes.}. 
However, such an outcome indicates the existence of destructive or constructive
interference among two or more amplitudes.  In practice, a sum of fit fractions
$\gg$ 100\%, that is
 a large destructive interference, is often indicative of
underlying problems in the fit model.
In the study under consideration it may be that the apparent large sum of
fractions is in fact an artifice, and that in the coherent sum the significant
NR contribution is largely "eliminated" through the destructive interference
arising from the $\sim 180^\circ$ phase difference between the two NR
amplitudes. 

Subsequently, several amplitude analyses, including 
Belle's study of the mode $B^0 \to K^0_S \pi^+\pi^-$~\cite{bellekspipi}
and the BABAR analysis of $B^0 \to K_S K^+ K^-$~\cite{babarkskk},
have used this exponential amplitude.
The fit results for share the same three
characteristics discussed above for $B^+ \to K^+ \pi^+ \pi^-$. 

This approach has been also used by both BABAR and Belle for three-body decays
involving two identical particles, namely
$B^+ \to K^+ K^+ K^-$ \cite{bellek2pi,babarkkk} and 
$B^+ \to \pi^+ \pi^+ \pi^-$ \cite{babarpipipi} \footnote{ 
In these cases the expression Eq.~(\ref{eq:anr}) contracts to a single
amplitude because of the symmetry under the exchange of identical particles.}.
Again, in both analyses a large contribution is found for the NR component,
and the overall sums of fractions are significantly above 100\%.

Another important problem for almost all the charmless three-body $B$ decays
is that it has been necessary to include some scalar contributions to get a good 
fit to data. The mass and width values obtained in some  analysis involving 
  $\pi^+ \pi^-$ in final state do not match among these two experiments  
\cite{bellek2pi,babark2pi} and some times do not match  between different channels 
of the same experiment \cite{babarpipipi}. The same situation occurs with high mass 
scalar contribution involving the final state $K^+ K^-$   \cite{bellek2pi,babarkkk}.

  The non-resonant component which, in general, spread within all the  
  populated phase space, can mimic other  dynamical components \cite{marina} through 
  the interferences with the  resonances present in the same phase space. 
High mass scalar resonances have also similar capability, due this angular 
distribution and the large width, these contribution  populates homogeneously large
portion of the  di-hadrons  phase space, without a clear signature like vector or tensor
contributions.  This   shadowing phenomenon was observed  in charm meson three body decays   
 at the  E791 experiment,  where the   overestimated contribution of the non-resonant 
amplitude use to be  able to mimic the real   contributions coming  from  the scalar mesons 
 $\sigma$ and $\kappa$\cite{sigma791,kappa791}.

\section{ Interference among resonances with the same quantum numbers.}

 In the isobar model, regarding interferences, all contributions are treated in the
same way, independent of its origin. On the other hand K-matrix approach proposed by FOCUS  
\cite{focus3pi,focusk2pi} use a especial treatment only to  represents the  S-wave scalar   
amplitude contribution  unitary  amplitude extract from two body elastic scattering. This approach  
take into account the peculiarity of long range phase space interference between two body amplitudes 
with the same  quantum numbers. The high spin resonances was treated as the isobar model.

     There are alternative   ways to write  two or more resonances with same quantum
numbers in a unitary fashion, Tornqvist \cite{torn}, per example,  
proposed to  use a  couple Breit-Wigner. There is a  more recent  representation
that can be very useful for three body amplitude analysis, it was 
proposed by Svec \cite{svec} based in a Hu method \cite{Hu}. As we can see below,
this approach can be useful, because it makes explicit the unitary 
 constraint in the isobar model, while  keeping its  usual structure. The proposal  
consist to represent a multi-resonance contribution through a unitary   product 
of isolated Breit-Wigner. Since  one  resonance is represented by an  unitary
S-matrix Breit-Wigner function the   product of Breit-Wigner will  be unitary also. 
This idea is suitable to Breit-Wigner or other amplitude to represent a resonance.

      With this assumption one can arrive to a simples T equation for two resonances
\cite{svec}:

\begin{eqnarray}
           T_{res} = BW_1(s) + BW_2(s) + 2 i BW_1(s)  BW_2(s) 
 \end{eqnarray}
        
where $BW_i$ is the usual Breit-Wigner function. This equation can be re-written as
coherent 
sum of Breit-Wigners with specific  complex coefficient, depending of the resonances
parameters 
as we can see below:

\begin{eqnarray}
         T_{res} = C_1^{(2)}(s)  BW_1(s) +  C_2^{(1)}(s)  BW_2(s) 
\end{eqnarray}

the coefficients are  given by $C_1^{(2)}(s) = 1-2i/(z_1-z_2)$ and 
$C_2^{(1)}(s) = 1+2i/(z_1-z_2)$, where  $z_i = m_i^2 - i m _1\Gamma_i(s)$
with a smooth dependence in the energy variable. So to guarantee unitarity, the phase differences between two
resonances with 
the same quantum number should be write as:

\begin{eqnarray} 
    C_1^{(2)}(s) - C_2^{(1)}(s)  = \frac{ 4 ( \Delta{m\Gamma(s)} - i \Delta{m^2})}{
\Delta^2{m\Gamma(s)} + \Delta^2{m}^2}
\end{eqnarray}

where m is the mass of the resonance, $\Gamma(s)$ the width,  $\Delta{m^2}$ and  
$\Delta{m\Gamma(s)}$ are respectively the mass  and the  $m\Gamma(s)$ differences between 
the two resonance. Since the $\Delta{m^2}$ is in general  bigger than the $\Delta{m\Gamma(s)}$, 
and the only dependence in {\it s} variable comes from the width, one can expect a little changes 
in the phase difference. In some sense, Equation 2 justify the use of coherent sum of Breit Wigner
 with constant parameters, at least for charm three body decay taking in 
account the reduct  phase space of this decay. With the  parameters extracted  from
data, one can check whether or not  these values are compatible with expected by a two body   
unitary behaviour of resonances in the same partial wave.   

The suitability  of the this approach  to represent data involving amplitudes with
the same quantum numbers,  could be observed in two occasions: 
the scalar amplitude and the phase motion obtained with the partial analysis
analysis (PWA), performed for the  
$D^+\to K^- \pi^+\pi^+$ \cite{brian,pwa_focus,pwa_cleo} and for $D^+_s\to \pi^-
\pi^+\pi^+$ \cite{antimo}  decays. Both decays 
were  well represented with a coherent sum of two scalar resonances contributions
with a phase difference, as it was plotted  in references  
\cite{antimo,brian,pwa_focus,pwa_cleo}.  However the K-matrix was able to represent
well only the  $D^+_s \to \pi^- \pi^+\pi^+$. The scalar 
$ K^- \pi^+$ amplitude and the phase behaviour of the $D^+\to K^- \pi^+\pi^+$ decay,
obtained with PWA differed substantially 
from the  $ K^- \pi^+$ elastic scattering experimental results \cite{LASS}. This
observed difference should be due  a strong re-scattering 
contribution in this three body final state \cite{pennington}, that can not be
represented in the K-matrix formalism  \cite{anisovich_sarantsev}.

\section{ Interference among resonances with  different quantum numbers.}

 We can divide the interference between resonances with different quantum numbers in
 Dalitz plot in two types:  
 parallel, i.e. defined at the same Dalitz variable  $s_{ij}$ and that can interfere in a
large portion of the phase space,  like  the vector resonance $\rho(770)$ and the
scalar one $\sigma(500)$ or with the  $f_0(980)$ in 
a decay involving two charged mesons $\pi$ in final state.
The  {\it parallel interference} have been described in several models through 
 local interferences for all phase space,  but due to the angular distribution, the 
phase space integral should  close to zero to parallel resonances with different spin. 
  
The second type we can call {\it crossing}, i.e. interference between resonances 
that cross each other in the phase space with different $s_{ij}$ Dalitz plot variables,  like  the 
interference between  the $\rho(770)$ resonance with the $K^*(890)$ in $B^+$ decay in $ K^+\pi^+ \pi^-$. 
The crossing resonances has the interference  in a limited  phase space region and the integral  
in general is different from  zero.  With a  coherent sum, this  term make a real contribution to the total amplitude square.   
Due to the formal complexity to work with this  contribution, some  Isobar approach to describe hadron hadron interaction, 
preferred to make the approximation that  interferences between crossing resonances
zero. This approximation was used in SLAC manual for three body partial analysis 
in the  eighties  \cite{ slac}. To performed the analysis for the final state $
K^-\pi^+\pi^- p$, they assume the integral of the crossing term between the $\rho$ and $K^*(890)$ is zero, beside
they can observe a significant amount of interference.

Since the crossing interference  can not be estimated by first principles, a correct inclusion of the crossing 
contribution in  amplitude analysis is a big challenger. The  best known formal approach to  use for study re-scattering
effects is  the Faddeev equation \cite{azimov, faddeev}. Very preliminary studies
using this method, have  presented nice results with a natural explanation for  the
observed phenomena to a global  phase difference between the P and S wave from $K
\pi$   partial wave analysis (PWA) of the   $ D^+ \to K^-\pi^+\pi^+$ decay to the $K \pi$
elastic scattering \cite{tobias}. In fact other than the difference  pointed out before between these
two experimental results, there  is a global  phase difference  bigger than $\pi/2$
degrees\cite{brian,pwa_focus,pwa_cleo}.  This difference between three body decay  and 
elastic scattering  was  also observed in $ D^+_s \to \pi^-\pi^+\pi^+$ partial wave analysis \cite{antimo}.

Experimental studies in the sixties and seventies presented a practical and
empirical approach for the 
re-scattering  problem. They treated  this effect by  a coherent sum of Breit Wigner
also to the crossing resonances. 
In  $K^+ p \to K^0 \pi^+ p $ studies at CERN experiment \cite{Jongejans2}, they  
introduced a sum of individual amplitudes with a constant phase difference, between
the interfering resonances due a re-scattering phenomena. This   determine the
degree of constructive or destructive interference among them. Each amplitude is
parametrized by a Breit-Wigner multiply by a real density term independent of the
energies, that can be extracted from data. However there is an important difference
between the original amplitude analysis and the one we use nowadays, the coherence
sum was not generalized and fixed for all resonances, like we have been doing in the
three body heavy mesons decays.

     In fact in their studies,  the coherence is include in the fit function only
among  resonances that they observe {\it a priory} in data, a signature of  interference
between them. 
For the  $K^+ p \to K^0\pi^+ p $ study, they use a coherent sum only for the 
$K^*(1400)p$ and a baryonic resonance amplitude $K^0N^*(1236)$, but not between the
dominant contributions $K^*(890)p$ and  $K^0N^*(1236)$. This approach was
generalized  by the Berkeley group that studied the  interaction 
$\pi^- p \to \pi^0\pi^+ p $ \cite{gidal}. In fact they introduced a coherence factor
to take in account the observed re-scattering for all resonances present in the
spectrum.  The observed results   between the $ \rho^- $'s amplitudes and the
different $N^*$ resonances variety from 0.6 to 0.8. 

Other physics process involving particle interference among different quantum
numbers used also a 
free parameter to manager with the non obvious source of coherence. In order to
observe the $\rho$ and 
$\omega$ isospin violation interaction proposed by Glashow \cite{Glashow},
experimental studies were performed with  low transverse momentum events from $\pi^-
p \to \pi^-\pi^+ n $ 
interaction \cite{SLAC_1972}.  The amplitude square used to fit the $\pi^+\pi^-$
spectrum,  included one Breit-Wigner square  for the   $\rho$ decays and another one
for the G-parity violating channel $\omega\to \pi^+\pi^-$, plus an interference
term. The last one was wrote to represent the clear interference observed in the
$\rho$ mass region. Besides  the usual product of two Breit-Wigners and a relative
strong phase, was also included a real multiplicative 
coherent factor that could change from  0 to 1. This factor was also include in
other studies $\rho$ and $\omega$
interactions studies with different reactions \cite{goldhaber,allison,hagopian}.

     More recently was proposed also a  coherence parameter to take in account possible contamination 
in the dominant pseudoscarlar  exchange process,  from the axial meson component in the   $\pi^- p \to \pi^0\pi^0 n
$ interaction \cite{achasov}. Also Cristal Barrel collaboration
\cite{cristal1,cristal2} used a coherence parameter to  take in account  the many
different partial waves in the initial and final $p \bar p$ interaction at rest,
that part interfere and part does not. For the 
  $p \bar p \to \pi^0 \eta\eta$ \cite{cristal2} they fit seven intermediary states,
with seven complex number for each one resonance contribution plus twelve coherence 
factor. Some of them were set to zero because there was no overlapping region or 
{\it apriori} tests did not present any significant  contribution.

     Heavy meson three body decays has a simpler  configuration since that it
involves an spin 0 initial state and three spin zero  final state. So the use of a coherence factor is less
crucial than used in the above examples, were some times  it is associated with and average of the
initial or final helicity state for interaction involving baryons. However some problems remain in heavy meson decays,
like the re-scattering, the  existence of different isospin amplitudes in a same final state. 

    Also a technical problem can coming out in amplitude analysis. In fact, experimental resolution have 
not been considered fully  in their analysis. In general  it is not important to describe the mass spectrum 
of the usual resonances, as it was pointed out for the narrow $f_0(980)$ resonance analysed in the 
$  D^+_s \to \pi^+\pi^-\pi^+ $ \cite{f0_e791}. However it might be different 
for describe the crossing region between two narrow  resonances, the interference
can  created a strong variation  in a short little  phase space region and the experimental 
resolution can spread out  this variations in this particular  region. One could take this effect 
in account in an amplitude fit function using Monte Carlo studies. But the 
inclusion of phenomenological coherence factor could  serve  to mimic this effect
and others possibles associated at the non-parametrized simple re-scattering contributions
 and the average on the isospin amplitudes.

\section{ Closing Remarks}

     Charmless three body B decays has an important appointment soon with the LHCb
data \cite{LHCb}. In fact for the LHC run 2010-2012  it is expect  more than decades of  
thousand events for the decays $B^+ $ to three light mesons and 
thousand of events to three body decays  channels involving a proton anti-proton  \cite{ICHEP2010}.  
This amount of events corresponds about  one order of magnitude bigger than  the produced by 
the electron positron factories Belle and BaBar until today. With this statistics  many new things  
could be understood for CP violation, that is the main proposal of LHCb  experiment. Model independent 
approach  was developed  to search new  sources of directly  CP violation in Dalitz plot \cite{mirandizing}. 
However if one want to go further and extract more information from Dalitz
plot, like the CKM phase $\gamma$ \cite{prl98,prd07} or the relative amount  from the Penguins and
Tree   contributions or also to know the relative CP violation from each intermediary resonance state, or
even to extract the the y parameter from $B^0_s$ \cite{zupan} one have to work with
amplitude analysis. But as we could see in this paper, many  studies have to be performed before to get  
trust-able parameters from this kind of fits. 

       We could see here that the first results from Belle and BaBar presented some
problems to describe the non-resonant contribution and the high mass scalar particles. The inclusion of
an empirical exponential function increased significantly the confidence level of the fit. However we noted
that this result imply in a strong destructive interference between the  $s_{13}$ and  $s_{23}$ non-resonant
components. This destructive interference take place just in the crossing region
between the  $\pi^+ \pi^-$ and   $K^+ \pi^-$ resonances. We think one of the first studies   to do 
with the new coming data is to look deeply at this effect. 
As it was pointed out in the previews section, the correct inclusion of the crossing
region is the big challenger and seems that the charmless B three body decays analysis have to 
take care of this problem in order to have a good representation of the big amount of events expected
for these final state B decays.  

       The start point of the amplitude analysis for these studies  could keep the
general form of the isobaric  model, but given more freedom to the fit function allowing more 
flexibility to understand better the data. Although Eq. 4 shown a smooth variation of the phase difference 
with data, this effect can be important for the fit involving a big amount of data as expected in 
LHCb and them could be add at the amplitude. In order to understand  the importance of the coherence 
in hadronic interaction, it would  be interesting  includes   a phenomenological real parameter in some 
 interference terms of the fit function. The coherence parameters values, obtained from fit 
can give some light for understand better how works interferences in the crossing
region among two resonances placed in different invariant mass, like  $\pi^+ \pi^-$
and   $K^+ \pi^-$   resonances in $B^+ \to  K^+ \pi^- \pi^+$ decay. 
   
{\bf Acknowledgments:} 
I would like to  thanks Jussara Miranda, Claúdio Lenz, Yakov Azimov and Alberto Reis  for many suggestions and 
clarifying conversations  about issues related with coherence in quantum mechanics.

\end{document}